\documentclass[letterpaper,english,pra, reprint, superscriptaddress, floatfix]{revtex4-1}
\usepackage[T1]{fontenc}
\usepackage[latin9]{inputenc}
\usepackage[pdftex]{color}
\usepackage{babel}
\usepackage{verbatim}
\usepackage{float}
\usepackage{amstext}
\usepackage{amssymb}
\usepackage[pdftex]{graphicx}
\usepackage{esint}
\usepackage[unicode=true,
 bookmarks=true,bookmarksnumbered=false,bookmarksopen=false,
 breaklinks=false,pdfborder={0 0 1},backref=false,colorlinks=true]
 {hyperref}
\hypersetup{pdftitle={Vibration-induced field fluctuations in a superconducting magnet },
 pdfauthor={J. W. Britton, B. C. Sawyer, J. G. Bohnet, H. Uys, M. J. Biercuk, J. J. Bollinger},
 pdfkeywords={superconductivity, superconducting magnet, electron spin resonance, ESR, NMR, ion cyclotron resonance, EPR, quantum logic, qubits}}

\makeatletter

\pdfpageheight\paperheight
\pdfpagewidth\paperwidth

\providecommand{\tabularnewline}{\\}

\pdfoutput=1

\newcommand{\ket}[1]{\left|#1\right>}

\newcommand{\nbep}[0]{^9\mbox{Be}^+}
\newcommand{\spinup}[0]{\ket{\uparrow}}
\newcommand{\spindn}[0]{\ket{\downarrow}}

\makeatother

\begin{document}

\title{Vibration-induced field fluctuations in a superconducting magnet }

\author{J. W. Britton}

\affiliation{Time and Frequency Division, National Institute of Standards and
Technology, Boulder, Colorado 80305, USA}

\affiliation{Army Research Lab, Adelphi, MD, 20783, USA}

\author{J. G. Bohnet}

\affiliation{Time and Frequency Division, National Institute of Standards and
Technology, Boulder, Colorado 80305, USA}

\author{B. C. Sawyer}

\affiliation{Georgia Tech Research Institute, Atlanta, Georgia 30332, USA}

\author{H. Uys}

\affiliation{Department of Physics, Stellenbosch University, 7600, Stellenbosch,
South Africa and National Laser Centre, Council for Scientific and
Industrial Research, Brummeria, 0184, Pretoria, South Africa}

\author{M. J. Biercuk}

\affiliation{ARC Center for Engineered Quantum Systems, School of Physics, The
University of Sydney, NSW 2006 Australia}

\author{J. J. Bollinger}

\affiliation{Time and Frequency Division, National Institute of Standards and
Technology, Boulder, Colorado 80305, USA}

\email{joe.britton@gmail.com}

\begin{abstract}
Superconducting magnets enable precise control of nuclear and electron
spins, and are used in experiments that explore biological and condensed
matter systems, and fundamental atomic particles. In high-precision
applications, a common view is that slow ($<1$~Hz) drift of the
homogeneous magnetic field limits control and measurement precision.
We report on previously undocumented higher-frequency field noise
($10$~Hz to $200$~Hz) that limits the coherence time of $\nbep$
electron-spin qubits in the $4.46$~T field of a superconducting
magnet. We measure a spin-echo $T_{2}$ coherence time of $\sim6\,\mbox{ms}$
for the $\nbep$ electron-spin resonance at $124\,\mbox{GHz}$, limited
by part-per-billion fractional fluctuations in the magnet's homogeneous
field. Vibration isolation of the magnet improved $T_{2}$ to $\sim50$~ms.
\end{abstract}
\maketitle
\maketitle

\section{INTRODUCTION}

In many spectroscopic applications the field stability of superconducing
magnets is important. For example, in atomic physics superconducting
magnets are used for high-precision mass spectroscopy \cite{vandyck_ultrastable_1999,myers_most_2013,rainville_ion_2004}
and stringent tests of quantum electrodynamics (QED) via magnetic
moment measurements \cite{odom_new_2006,hanneke_new_2008,sturm_g_2011,sturm_highprecision_2014}.
These many-hour experiments typically involve repeated quantum state
preparation, evolution and measurement. Slow drift of the homogeneous
magnetic field is commonly believed to limit accurate comparison of
sequential measurements.

A variety of techniques can improve long-term field stability to as
low as $2\times10^{-11}$/hr, enabling ion cyclotron mass spectroscopy
at the $\sim10^{-10}$ level \cite{vandyck_ultrastable_1999} or better.
One source of drift is the temperature-dependent susceptibility of
in-bore materials \cite{salinger_magnetic_1961,xia_magnetization_2002};
controlling liquid-cryogen boil-off rate (which varies with atmospheric
pressure) and stabilization of the cryostat exterior temperature improves
stability by reducing time-variation of thermal gradients \cite{vandyck_ultrastable_1999}.
Sensitivity to ambient laboratory fields can be passively reduced
by self-shielding \cite{gabrielse_selfshielding_1988} and by active
feedback (to $10$~Hz) \cite{vandyck_ultrastable_1999,myers_most_2013}.
Higher frequency fluctuations induced by vibrations and their mitigation
are not well documented in the literature. Although high-field NMR
installations ($>500$~MHz proton frequency) routinely employ vibration
isolation \cite{schneider-mantau1997high_field_magnets,kiyoshi2010hts_nmr},
the fast part-per-billion noise we document here is beyond the detection
capability of NMR, requiring high-frequency ESR (see Appendix).

\section{APPARATUS}

In the experiments described here, the homogeneous field of a superconducting
magnet gives radial confinement of $\nbep$ ions in a Penning trap,
and defines the quantizing axis for the $\nbep$ ions' valence-electron
spin degree of freedom. Electron spin resonance (ESR) with these spins
is sensitive to part-per-billion fractional field fluctuations at
frequencies out to several hundred hertz (see Appendix). Our measurements
show correlations between the $T_{2}$ coherence of the electronic
spin states with measured mechanical vibration of the magnet dewar.
We further demonstrate that spin coherence may be extended by acoustic
and vibration isolation of the magnet system.

\begin{figure}
\includegraphics[width=1\columnwidth]{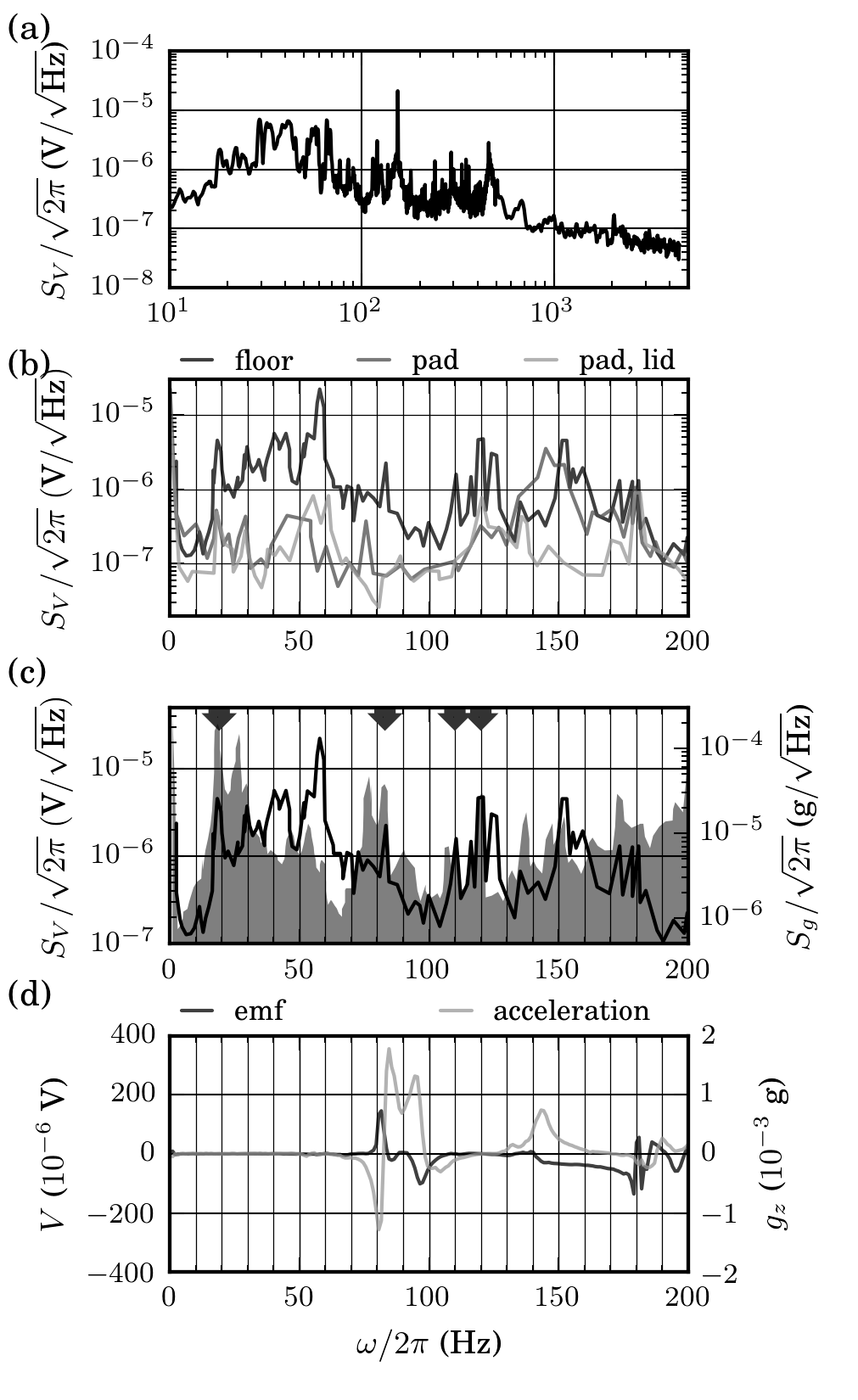}

\caption{An electromotive potential $V_{emf}(t)$ is induced on the magnet's
normal $Z_{0}$ shim coil by fluctuating magnetic fields in the magnet
bore. The voltage spectral density ($S_{V}$) of $V_{emf}$ is plotted.
(a) Representative data from 2008. For $\omega/2\pi>100$~Hz, $S_{V}$
scales as $\omega^{-1}$ \cite{Biercuk2009}. (b,c) For the range
$\omega/2\pi\lesssim200$~Hz we observe that $S_{V}$ is correlated
with some (but not all) ambient seismic and acoustic noise in the
lab. (b) $S_{V}$ for the magnet coupled to the laboratory floor (black
line). Several spectral features are suppressed by supporting the
magnet on rubber vibration-isolation pads (dark gray line). The broad
noise peak at $155$~Hz is suppressed by covering the top side of
the magnet bore with a lab notebook (light gray line). (c) The acceleration
spectral density in the z-direction ($S_{g}$, grey shaded area) and
the $S_{V}$ of $V_{emf}$ (black line) are plotted for case of the
magnet coupled to the laboratory floor. Correlations in the power
spectra are marked with black arrows. (d) Lock-in detection at frequency
$\omega/2\pi$ for signals $V_{emf}$ (black line) and acceleration
$g_{z}$ (gray line) in response to a mechanical oscillator on top
of the magnet dewar oscillating at frequency $\omega$. A differential
relationship is expected between $V$ and $g_{z}$; a peak in the
emf is expected at a zero crossing in $g_{z}$ as at 83~Hz. The strongest
correlation between resonant (d) and ambient (c) response is 83~Hz.
\label{fig:ambient_emf_gz}\label{fig:thumper}}
\end{figure}

Penning traps are routinely used in a wide range of studies including
mass spectrometry of biological molecules (e.g.,~FTICR) \cite{marshall_fourier_1998}
and tests of fundamental physics by precision spectroscopy (e.g.,~QED)
\cite{odom_new_2006,hanneke_new_2008,sturm_g_2011,myers_most_2013,sturm_highprecision_2014}.
Research aims of the NIST Penning trap include simulation of quantum
magnetism \cite{britton_engineered_2012} and potentially quantum
computation \cite{Porras2006,baltrusch_fast_2011}. Details of our
setup have been described previously \cite{britton_engineered_2012,biercuk_highfidelity_2009}.
Here we summarize features important for measuring the magnetic field
noise and emphasize relevant system modifications since Ref.~\cite{biercuk_highfidelity_2009}. 

We confine a laser-cooled crystal of $N\sim300$ $\nbep$ ions in
a Penning trap with a $B_{0}=4.46$~T superconducting magnet. Our
two-level system (qubit) is the $\nbep$ valence electron spin states
$\ket{\uparrow}\equiv\ket{m_{s}=+1/2}$ and $\ket{\downarrow}\equiv\ket{m_{s}=-1/2}$,
where $m_{s}$ is the spin's projection along the $B_{0}\hat{z}$
quantizing field. The spins' Larmor precession frequency $\Omega_{0}$
is first-order field sensitive: $\hbar\Omega_{0}\simeq g_{J}\mu_{B}B_{0}$,
where $g_{J}\simeq-2.002$ is the electron g-factor and $\mu_{B}$
is the Bohr magneton. Including a small hyperfine correction, $\Omega_{o}=2\pi\times124.05\,\mbox{GHz}$
for $B_{0}=4.46$~T. Similar to other nuclear magnetic resonance
(NMR) and electron spin resonance (ESR) experiments, arbitrary spin
rotations are obtained with a resonant external microwave-frequency
magnetic field $B_{\mbox{\text{rf}}}\cos(\Omega_{0}t)$. The microwave
field is approximately uniform across all the ions and drives a spin
flip $\tau_{\pi}=68\,\mu\mbox{s}$. 

The superconducting magnet used in these experiments is a room-temperature
bore ($12.7\,\mbox{cm}$~diameter) Nalorac Cryogenics Corporation\cite{nist_nist}
model 4.5/125 manufactured in 1990 \cite{nist_nist}. The $\sim50\,\mbox{cm}$
tall NbTi superconducting solenoid produces a $4.5$~T field for
a continuous current of $58\,\mbox{A}$. Normal $Z_{0}$, $Z_{1}$
and $Z_{2}$ shim coils are wound on the room temperature bore tube.
The main superconducting coil and 8 superconducting shim coils are
fully immersed in a reservoir of liquid helium; the liquid helium
boil-off rate is 20~mL/hour. A liquid nitrogen sheath bears the brunt
of the thermal load not deflected by layers of super-insulation. Nested
vessels containing the liquid cryogens hang from attachment points
within the cryogen-fill towers. Dielectric struts near the bottom
of the dewar hold vessels apart and presumably provide some degree
of mechanical damping. The dewar is constructed almost entirely of
aluminum. The magnet's circulating super-current was stable from 2004-2013
and 2014-present. The vacuum envelope of the Penning trap is rigidly
attached to the magnet dewar. 

\begin{figure}[!t]
\includegraphics[width=1\columnwidth]{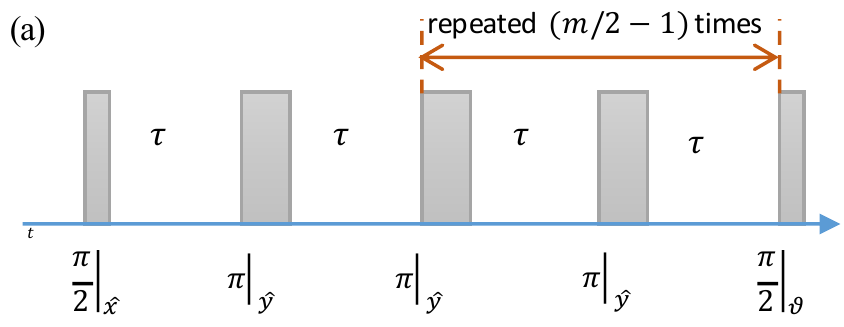}
\includegraphics[width=1\columnwidth]{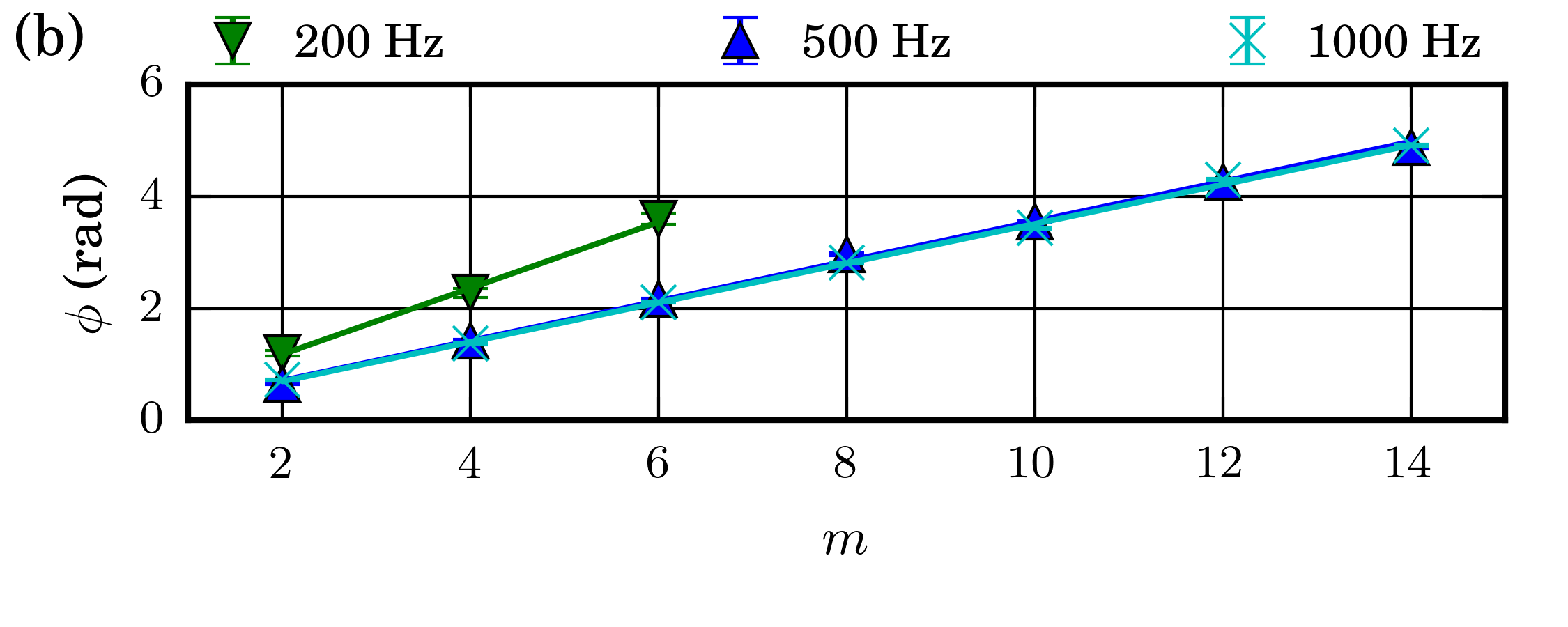}

\caption{\textbf{(a)} We measure the $T_{2}$ coherence of our spins using
a spin-echo sequence ($m=2$).\textbf{ }The spin-echo is first-order
sensitive to coherent, synchronous field oscillations at frequency
$\omega$ when $\tau=\frac{1}{2}(\omega/2\pi)^{-1}$. For $m=2$ (interval
$2\tau$) phase $2\phi$ is acquired by the ion spins, where $\phi\propto\int_{0}^{\tau}B(t)dt$.
\textbf{(b)} An independent calibration of the sensitivity of the
$Z_{0}$ shim coil using $m$ spin-echo sequences (see main text).
We observe $d\phi/dm$ is linear for EMF-induced magnetic fields at
$\omega/2\pi=200\,\mbox{Hz}$, $\omega/2\pi=500\,\mbox{Hz}$ and $\omega/2\pi=1000\,\mbox{Hz}$.
\label{fig:eta_calibration} \label{fig:cpmg_sequence} }
\end{figure}

\begin{figure*}[!t]
\includegraphics{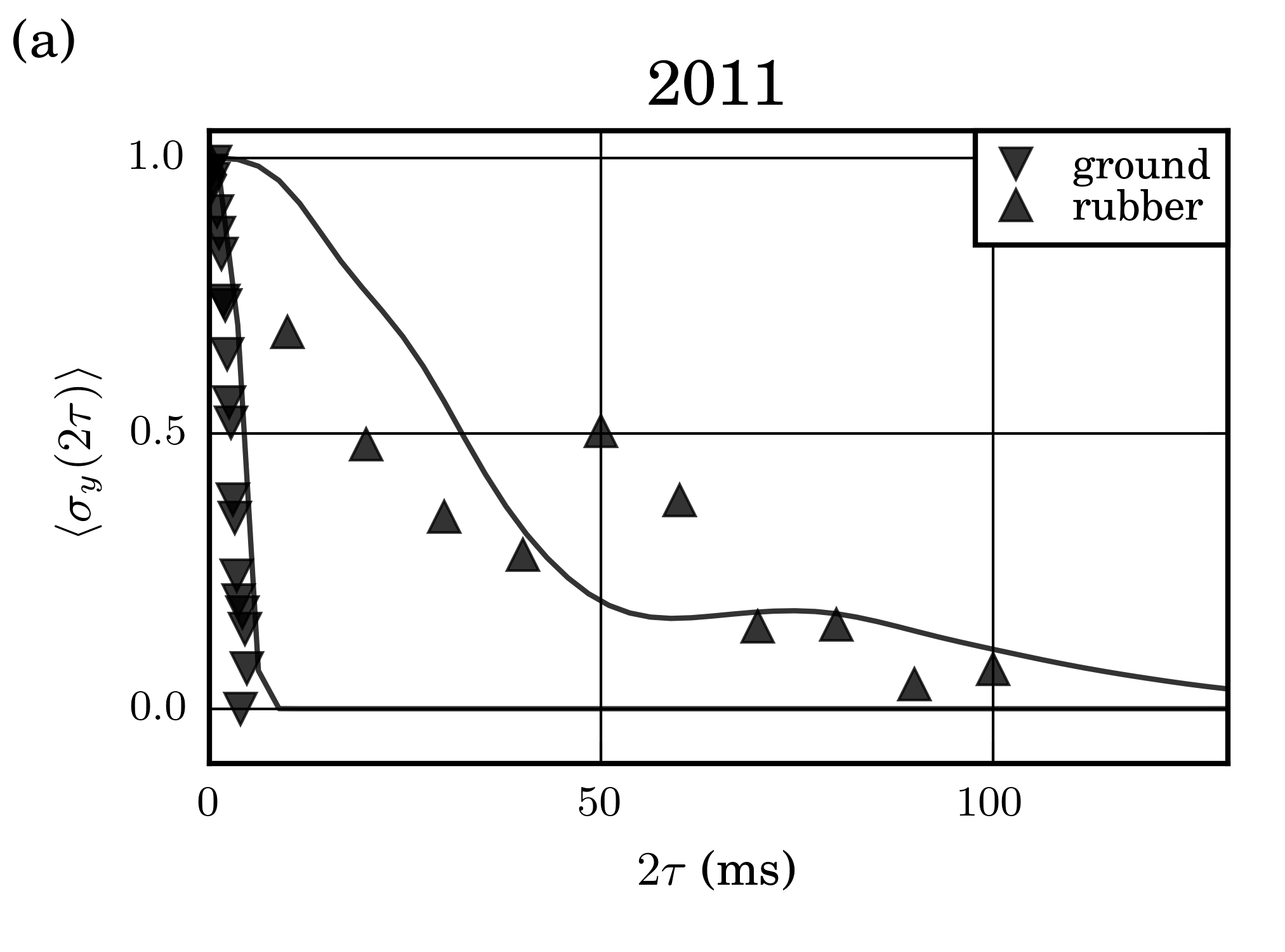}\includegraphics{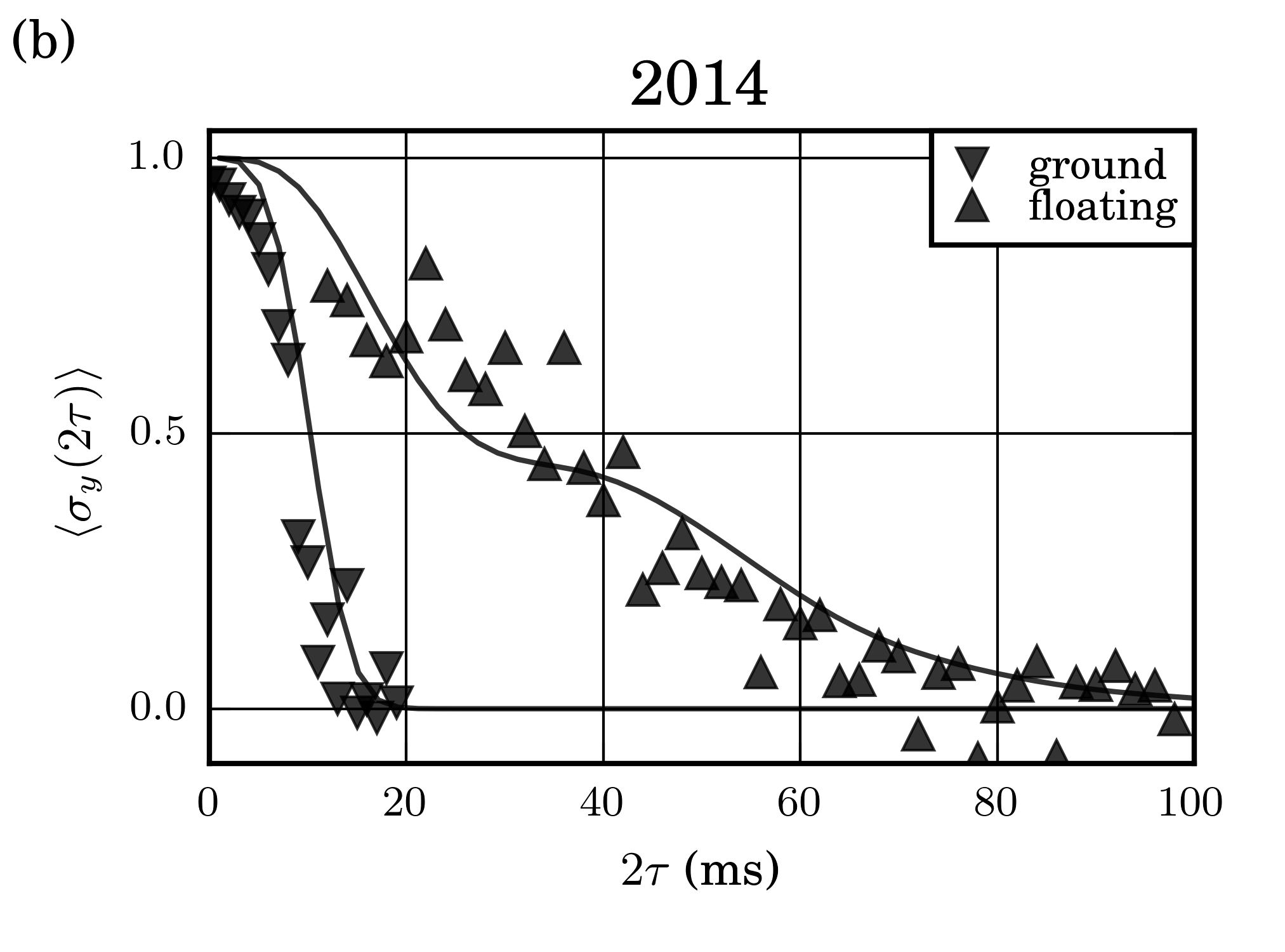}

\caption{Isolation of the magnet dewar from laboratory floor vibration increases
$T_{2}$. (a, b) Plots of spin Bloch vector length after a single
$\pi$-pulse spin echo experiment ($m=2$) with total evolution time
$2\tau$ (points). Solid lines are theory curves given a magnetic
field noise spectrum inferred from $V_{emf}$ with a single, frequency-independent
$\eta$ determined by best fit. (a) In 2011 we contrast $T_{2}$ when
the magnet is in contact with floor vibration vs isolated by rubber
isolation pads; a fit yields $\eta=15$~$\mbox{m}^{2}$. b) A similar
experiment was performed in $2014$ but with the magnet supported
by a floating optics table; a fit yields $\eta=11$~$\mbox{m}^{2}$. }
\label{fig:T2}
\end{figure*}

\section{MEASUREMENTS}

In 2010 we noticed a correlation between mechanical vibration of lab
floor and the potential across our magnet's normal-current in-bore
$Z_{0}$ shim coil $V_{emf}(t)$. For the ambient laboratory environment,
Figures~\ref{fig:ambient_emf_gz}a-c show $V_{emf}$ as a voltage
spectral density ($S_{V}$, $V/\sqrt{\mbox{s}^{-1}}$) and Figure~\ref{fig:ambient_emf_gz}c
shows the vibration measured as an acceleration spectral density ($S_{g}$,
$g/\sqrt{\mbox{s}^{-1}}$). Acoustics also contribute to $V_{emf}$.
The fundamental for a hollow pipe is $f_{open}=v(2L+1.6d)^{-1}=155\,\mbox{Hz}$,
where $v=343\,\mbox{m/s}$ is the velocity of sound in air, and for
our magnet bore, $L=1\,\mbox{m}$ and $d=0.127\,\mbox{m}$. With an
open magnet bore we see a broad spectral feature at $\sim150$~Hz
that is strongly attenuated when the bore is covered by a lab notebook
(Fig.~\ref{fig:ambient_emf_gz}b). Mechanical resonances in the dewar
likely contribute to $V_{emf}$ as well. To explore this we induced
mechanical motion of the magnet along the solenoid axis by applying
a coherent driving force at frequency $\omega/2\pi$ to the top of
the magnet dewar using an electromechanical oscillator (EMO) rigidly
mounted to the top of the dewar. The EMO is a speaker solenoid (no
diaphragm) driven by a sinusoid. We use lock-in detection to measure
$V_{emf}(\omega)$ induced by the EMO drive. Resonances are observed
in the frequency range of interest (Fig.~\ref{fig:ambient_emf_gz}d)
which hints at complex electromechanical couplings. Despite this complexity,
we anticipated $S_{V}$ could be indicative of magnet field fluctuations
and $T_{2}$ coherence. 

We measure the $T_{2}$ coherence of our spins using a spin-echo sequence
illustrated in Figure~\ref{fig:cpmg_sequence} with a single $\pi$-pulse
($m=2$). Optical pumping to $\ket{\uparrow}$ followed by a $\pi/2$-pulse
rotates the spins to the equatorial plane of the Bloch sphere, defined
to be the $\hat{y}$-axis in the rotating frame of the applied microwaves.
At the end of the spin-echo evolution interval $2\tau$, we apply
a final $\pi/2$-pulse with a phase shift $\theta$ relative the initial
$\pi/2$-pulse, and then make a projective measurement of the z-component
of the spins (see Appendix). For a given $\theta$ we repeat this
experiment many times. From the phase and contrast of the resulting
fringe pattern (obtained by varying $\theta$), we measure that on
average the spins maintain their alignment with $\hat{y}$-axis but
with reduced coherence $\left<\vec{\sigma}(2\tau)\right>=\left<\hat{\sigma}_{y}(2\tau)\right>$.
$T_{2}$ corresponds to $T_{2}=2\tau$ when $\left<\hat{\sigma}_{y}(2\tau)\right>=e^{-1}$.
We note that our measurement of $T_{2}$ is sensitive to the phase
evolution of the spins relative to the applied microwaves and is therefore
sensitive to magnetic field fluctuations. This may be contrasted with
many NMR and ESR experiments that determine the magnitude of the transverse
spin coherence through measurement of both spin quadratures (see,
for example, \cite{morley_multifrequency_2008}). The spins' longitudinal
relaxation $T_{1}$ is effectively infinite. We observed reduced $T_{2}$
under conditions when $S_{V}$ is large (Fig.~\ref{fig:T2}). 

Our key observations were consistently observed over an interval of
5 years in two different laboratory environments using two Penning
traps. In 2011 (Fig.~\ref{fig:T2}a) we observed an order of magnitude
improvement in $T_{2}$ upon isolating the superconducting magnet,
optics and Penning trap from the laboratory floor using rubber flexural
mounts ($\sim7$~Hz resonant frequency, Barry Controls p/n 633A-260)\cite{nist_nist}.
Encouraged by this result, we moved our apparatus to a new lab space
in $2014$ which has lower ambient acoustic and seismic noise. In
this new lab with the magnet coupled to the floor we observe increased
$T_{2}$ (Fig.~\ref{fig:T2}b) relative to the old lab under the
same conditions. In the new lab we isolated the magnet, trap and optics
from seismic vibration using pneumatic legs that support our optics
table ($\sim1$~Hz resonant frequency). The impact of magnet vibration
isolation in both labs is an order of magnitude improvement in $T_{2}$
(Fig.~\ref{fig:T2}). Despite the lower noise levels in the new lab,
the $T_{2}$ observed with the magnet isolated is about the same as
in the old lab. In the Appendix the potential sensitivity of $T_{2}$
to other factors is discussed including 124~GHz phase noise, magnetic
field gradients, and the presence of permeable materials in the magnet
bore. 

The observed increase in $T_{2}$ can be causally related to $V_{emf}$
by noting that for a coil, a time-varying magnetic field $B(t)=B\sin(\omega t)$
induces a potential $V_{emf}(t)=V\cos(\omega t)$ with relative amplitude
$V/B=\eta\omega$. Here, $\eta$ is an unknown geometric factor with
units of $\mbox{m}^{2}$. We assume that $\eta$ is frequency-invariant
$S_{B}(\omega)=\frac{S_{V}(\omega)}{\eta\omega}$, where $S_{B}(\omega)$
is the magnetic field spectral density in T/$\sqrt{\mbox{s}^{-1}}$.
We use a filter function formalism to predict the spin coherence \cite{Biercuk2009,biercuk_dynamical_2011}.
For a spin initially oriented along the y-axis, its average projection
along $\hat{y}$ after interval $2\tau$ is 
\begin{equation}
\left<\sigma_{y}(2\tau)\right>=e^{-\chi(2\tau)},
\end{equation}
where 
\begin{equation}
\chi(2\tau)=\frac{1}{2\pi}\int_{0}^{\infty}S_{\beta}^{2}(\omega)F_{1}(\omega,2\tau)\omega^{-2}d\omega.\label{eq_chi_coh_decay}
\end{equation}
Here, $S_{\beta}(\omega)=S_{B}(\omega)g\mu_{B}/\hbar$ is the spectral
density of frequency fluctuations in units of $\mbox{s}^{-1}/\sqrt{\text{s}^{-1}}$
and $F_{1}(\omega,2\tau)=16(\sin(\omega2\tau/4))^{4}$ is the spin-echo
filter function in the limit of zero pi-pulse length. This model is
in qualitative agreement with observed decay of coherence in Figure~\ref{fig:T2}
taking $\eta$ to be a free parameter. From these fits we obtain $\eta=15$~$\mbox{m}^{2}$
($2011$) and $\eta=11$~$\mbox{m}^{2}$ ($2014)$. 

An independent calibration of $\eta$ was obtained by coherent shaking
of the magnet at $\omega$ using the EMO and observing $V_{emf}(\omega)$
and $B(\omega)$ using a spectrum analyzer and the ion-spins respectively.
The EMO-induced motion induces a magnetic field \textbf{$B(t)$ }that
gives rise to a potential $V_{0}\cos(\omega t)$ across the $Z_{0}$
shim coil, where $B(t)=V_{0}\sin(\omega t)/\omega\eta$. Synchronous
with the EMO drive, we perform a pulse sequence with $m-1$ equally
spaced $\pi$-pulses and $m$ intervals of duration $\tau=\frac{1}{2}(\omega/2\pi)^{-1}$.
This sequence (Fig~\ref{fig:eta_calibration}a) is sensitive to a
time-varying field at $\omega$. For $m=2,$ during interval $2\tau$
a phase $2\phi$ is acquired by the ion spins, where $\phi=\int_{0}^{\tau}\Delta(t)dt$
and $\Delta(t)=g\mu_{B}B(t)/\hbar$ is the shift in Larmor frequency
due to $B(t)$. Each $\tau$ includes the $68\,\mu\mbox{s}$ $\pi$-pulse
time. The acquired phase $\phi$ is measured by varying the final
$\pi/2$-pulse phase $\theta$ and extracting the phase of the resulting
sinusoidal fringe pattern. A pulse sequence of length $m$ yields
phase accrual $m\phi$ (Fig~\ref{fig:eta_calibration}b). From the
slope $d\phi/dm$ and $V_{0}$, we calculate $\eta=2g\mu_{B}V_{0}/(\hbar\omega^{2}\frac{d\phi}{dm})$.
This process was repeated for several EMO drive frequencies. For each
$\omega$ the EMO drive phase was adjusted to maximize $\phi$. We
observed $\eta=37$~$\mbox{m}^{2}$ at $200$~Hz, $\eta=7$~$\mbox{m}^{2}$
at $500$~Hz and $\eta=12$~$\mbox{m}^{2}$ at $1000$~Hz. The
dependence of $\eta$ on frequency indicates that our assumption of
frequency independence of $\eta$ in the fits of Fig.~\ref{fig:T2}
is simplistic. Nevertheless the mean values of $\eta$ obtained from
Fig.~\ref{fig:T2} and $d\phi/dm$ are similar. A better understanding
of $\eta(\omega)$ is hampered by our poor knowledge of the exact
geometry and mechanical support of the coils in our  magnet. 

The RMS variation in the spin-flip frequency $\delta_{RMS}$ due to
$S_{B}$ is 
\begin{equation}
\delta_{RMS}^{2}=\frac{1}{\pi}\int_{0}^{\infty}S_{\beta}^{2}(\omega)d\omega
\end{equation}
(see Appendix). Integrating $S_{\beta}^{2}$ for the conditions in
Figure~\ref{fig:T2} yields $\delta_{RMS}/2\pi$ equal to $135$~Hz
($1.1$~ppb) and 12~Hz ($0.1$~ppb) for the 2011 $S_{V}$ data
and 68~Hz ($0.5$~ppb) and 14~Hz (0.1~ppb) for the 2014 $S_{V}$
data. To the the best of our knowledge this is the first observation
of $0.1$~ppb short-term stability in the electron spin-flip frequency
stability in a high-field superconducting magnet.

\section{CONCLUSIONS}

In summary, we observed part-per-billion fractional fluctuations in
the homogeneous magnetic field of a $4.46$~T superconducting magnet
at frequencies up to $200$~Hz.  Using $\nbep$ electron-spins as
sensitive field detectors, an order of magnitude reduction in integrated
magnetic field noise was obtained by isolating the magnet from environmental
acoustic and mechanical noise. In so much as our superconducting magnet
is representative, we anticipate that a variety of high-precision
measurements may be limited by similar fluctuations and could benefit
from improved isolation. Examples include ESR and electron-cyclotron
resonance with frequencies $\geq90$~GHz (see Appendix) \cite{odom_new_2006,hanneke_new_2008,sturm_g_2011,sturm_highprecision_2014,ahokas_clock_2008,takahashi_quenching_2008,morley_longlived_2008}.
Fast field noise may underlie the $1\times10^{-9}$ line-shape broadening
observed but not well understood in single-electron g-factor experiments
($5.4$~T)~\cite{odom_new_2006,hanneke_new_2008} and in bound-electron
magnetic moment measurements of hydrogenic $^{28}$Si$^{13+}$ and
$^{12}$C$^{5+}$ (\textbf{$3.8$}~T)~\cite{sturm_g_2011,sturm_highprecision_2014}.
High-frequency field fluctuations may also be important in high-field
solid state ESR experiments where some of the longest reported T2
coherence times are a few hundred microseconds (8.5~T)~\cite{takahashi_quenching_2008,morley_longlived_2008}.
New materials with intrinsically longer spin relaxation times are
in development \cite{tyryshkin_electron_2011} and will require attention
to field noise in the regime we discuss in this paper. The well-defined
phase relationship between instantaneous $V_{emf}$ and magnetic field,
suggests spin coherence could be further increased by feeding forward
on the microwave phase.

\section*{ACKNOWLEDGMENTS}

We acknowledge financial support from DARPA OLE and NIST. The authors
thank Stephen Lyon and Gavin Morley for very useful discussions on
solid-state ESR experiments, and Joshua Savory and Stephen Russek
for their comments on the manuscript. This work includes contributions
by NIST and as such is not subject to U.S. copyright.

\section*{APPENDIX}

\subsection*{1. Penning trap}

A Penning trap relies on static magnetic and electric fields to achieve
3-dimensional confinement of ions. In equilibrium, the ion crystal
rotates at angular frequency $\omega_{r}$ (about $\hat{z}$) and
the Lorentz force ($q\vec{v}\times\mbox{\ensuremath{\vec{B}}}$) provides
a radial restoring potential in the strong, homogeneous magnetic field
$B_{0}\hat{z}$ (here, $B_{0}=4.46$~T). A static quadrupole electric
potential gives axial trapping (along $\hat{z}$). The trap potential
in a frame rotating at $\omega_{r}$ is 
\begin{equation}
q\phi(r,z)=\frac{1}{2}M\omega_{z}^{2}(z^{2}+\beta_{r}r^{2}),\label{eq:trapPotential-1}
\end{equation}
where $q$ is the ion charge, $M$ is the single-ion mass, $\beta_{r}=\omega_{r}\omega_{z}^{-2}(\Omega_{c}-\omega_{r})-1/2$,
$\Omega_{c}$ is the single-ion cyclotron frequency and $\omega_{z}$
is ions' harmonic center-of-mass motion along $\hat{z}$. For $\nbep$
ions in our trap potentials, $\Omega_{c}=B_{0}q/M=2\pi\times7.6$~MHz
and $\omega_{\mbox{z}}\sim2\pi\times800$~kHz. Ion rotation is precisely
controlled with an external rotating quadrupole potential \cite{mitchell_direct_1998}.
We set $\omega_{r}\sim2\pi\times45$~kHz so that the radial confinement
is weak compared to the axial confinement ($\beta_{r}\ll1$). Upon
Doppler laser cooling the ions' motional degrees of freedom ($T\sim1$~mK)\cite{sawyer_spectroscopy_2012},
the ions naturally form a 2D Coulomb crystal consisting of 1-4 planes
of ions. For $N\sim300$ ions, the crystal diameter is $\leq500\,\mu\mbox{m}$,
and $\leq60\,\mu\mbox{m}$ along the magnetic field. The separation
between planes is $\sim20\,\mu\mbox{m}$.

Quantum control experiments begin with Doppler laser cooling followed
by optical pumping to the $\spinup$ state using $\sim313\,\mbox{nm}$
laser light \cite{biercuk_highfidelity_2009}. Projective readout
of the ions' spin state is obtained by illuminating the ion crystal
with a laser beam tuned to a cycling transition resonant with the
$\spinup$ state and collecting fluorescence on a PMT; $\spinup$
ions appear bright, $\spindn$ ions appear dark. State preparation
and detection requires $5$~ms. Typical short-time Rabi flopping
traces exhibit $>99\%$ contrast. 

In the limit of large magnetic field, the $\nbep$ nuclear spin is
decoupled from the single valence electron. Optical pumping prepares
the nuclear spin in the $\left|m_{I}=+3/2\right>$ state.

\subsection*{2. Impact of magnetic field noise: ESR vs NMR}

In high-field magnetic fields the impact of small, fast magnetic field
fluctuations is qualitatively different for ESR than for NMR and ion-cyclotron
mass spectroscopy. Consider the case of a static magnetic field $B_{0}$
modulated at a single frequency $\omega_{m}$ with amplitude $B_{m}$, 

\begin{equation}
B(t)=B_{o}+B_{m}\sin\left(\omega_{m}t\right).\label{eq_B_modulation}
\end{equation}
The instantaneous Larmor precession frequency is 
\begin{equation}
\omega(t)=\Omega_{o}+\delta\omega_{m}\sin(\omega_{m}t),\label{eq_f_modulation}
\end{equation}
where $\delta\omega\equiv\gamma B_{0}$, where $\gamma$ is the gyromagnetic
ratio. The value of $\gamma$ is $g_{n}\mu_{N}$ for a nuclear spin
and $g_{e}\mu_{B}$ for an electron spin, where $g_{e}$($g_{n}$)
is the electron (nuclear) g-factor and $\mu_{B}$($\mu_{N}$) is the
Bohr (nuclear) magneton. The frequency modulation of Eq.~\ref{eq_f_modulation}
produces phase modulation of depth $\beta_{m}\equiv\delta\omega_{m}/\omega_{m}$
that results in sidebands at $\Omega_{0}\pm n\omega_{m}$ of relative
strength $J_{n}(\beta_{m})$ and a depleted carrier of relative strength
$J_{0}(\beta_{m})$ where 
\begin{equation}
\beta_{m}\equiv\frac{\delta\omega_{m}}{\omega_{m}}=\frac{\frac{B_{m}}{B_{0}}\Omega_{0}}{\omega_{m}}=\frac{B_{m}}{B_{0}}\frac{\Omega_{0}}{\omega_{m}}.\label{eq_modulation_strength}
\end{equation}
Since $J_{0}(\beta_{m})\sim1-\frac{1}{4}\beta_{m}^{2}$, the carrier
is substantially depleted for $\beta_{m}\sim2$. Broadband noise also
causes carrier attenuation. 

The impact of Eq.~\ref{eq_B_modulation} on ESR and NMR is strikingly
different ( Table~\ref{table_ESRvsNMR}). Suppose the field used
in the present ESR experiment was subject to ppb fractional fluctuations
$B_{m}/B_{0}=1\times10^{-9}$ at $\omega_{m}/2\pi=50$~Hz on top
of the homogeneous field $B_{0}$ responsible for the $\Omega_{0}/2\pi=124\times10^{9}$~Hz
Larmor precession (carrier). In this case, $\beta_{m}=2.5$ and the
carrier is fully depleted. In the case of a nuclear spin (NMR), the
fractional sensitivity is reduced by $\sim\mu_{B}/\mu_{N}=1836$ ($\mu_{N}$
is the nuclear magneton), $\beta_{m}=1\times10^{-3}$, and the carrier
suffers no depletion. 

A complimentary analysis in the main text relates the power spectral
density of frequency fluctuations $S_{\beta}^{2}(\omega)$ to the
spin coherence (Eq.~\ref{eq_chi_coh_decay}). The quadratic dependence
of $S_{\beta}^{2}(\omega)$ on $\gamma$ means decoherence is about
a million times weaker for NMR than for ESR 
\[
\chi_{NMR}/\chi_{ESR}=(\mu_{N}/\mu_{B})^{2}=3\times10^{-7}.
\]
That is, field noise that fully depletes (dephases) the Larmor carrier
in the context of ESR has negligible impact on NMR. 

\begin{table}[H]
\begin{centering}
\begin{tabular}{|c|c|c|c|}
\hline 
 & $\beta_{m}$ & $J_{0}$ & $J_{1}$\tabularnewline
\hline 
\hline 
ESR & 2.5 & -0.04 & 0.49\tabularnewline
\hline 
NMR & $1\times10^{-3}$ & 1 & $7\times10^{-4}$\tabularnewline
\hline 
\end{tabular}
\par\end{centering}

\caption{Equation~\ref{eq_modulation_strength} expresses the difference in
sensitivity of NMR and ESR experiments. Suppose the field used in
the present ESR experiment ($\Omega_{0}/2\pi=124\times10^{9}$~Hz)
was subject to ppb fractional fluctuations $\delta B/B_{0}=1\times10^{-9}$
at $\omega_{m}/2\pi=50$~Hz. This corresponds to $\beta_{m}=2.5$.
For a nuclear spin (NMR) the equivalent modulation index is $\beta_{m}=1\times10^{-3}$.}

\label{table_ESRvsNMR}
\end{table}

\subsection*{3. RMS Magnetic Field Variation}

We calculate the RMS variation in spin-flip frequency $\delta_{RMS}$
due to $S_{B}$ as follows.  Consider a Ramsey experiment with a
free-evolution interval $\tau$ short enough that the spin-flip frequency
remains constant over $\tau$. Let $\delta$ be the instantaneous
frequency deviation of the spin-flip frequency from its mean. The
accumulated phase difference between the microwaves and the spins
is then $\phi=\delta\tau$. In the limit $\delta_{RMS}\tau\ll1$ and
averaging over many experiment repetitions, the Bloch vector length
is 
\begin{equation}
\left<\hat{\sigma}_{y}\right>=\left<\cos\phi\right>\simeq1-\frac{1}{2}\delta_{RMS}^{2}\tau^{2}.
\end{equation}
Applying the filter function formalism and expanding to first order, 

\begin{equation}
\begin{array}{ccl}
e^{-\chi(\tau)} & \simeq & 1-\chi(\tau)\\
 & = & 1-\frac{1}{2\pi}\int_{0}^{\infty}S_{\beta}^{2}(\omega)F_{0}(\omega,\tau)\omega^{-2}d\omega\\
 & = & 1-\{\frac{1}{2\pi}\int_{0}^{\infty}S_{\beta}^{2}(\omega)d\omega\}\tau^{2},
\end{array}
\end{equation}
where $F_{0}(\omega,\tau)=4\sin^{2}(\omega\tau/2)$ is the Ramsey
filter function and we have used the small angle approximation. We
then obtain 
\begin{equation}
\delta_{RMS}^{2}=\frac{1}{\pi}\int_{0}^{\infty}S_{\beta}^{2}(\omega)d\omega.
\end{equation}

\subsection*{4. Microwave phase noise}

\begin{figure}[!h]
\includegraphics[width=0.7\columnwidth]{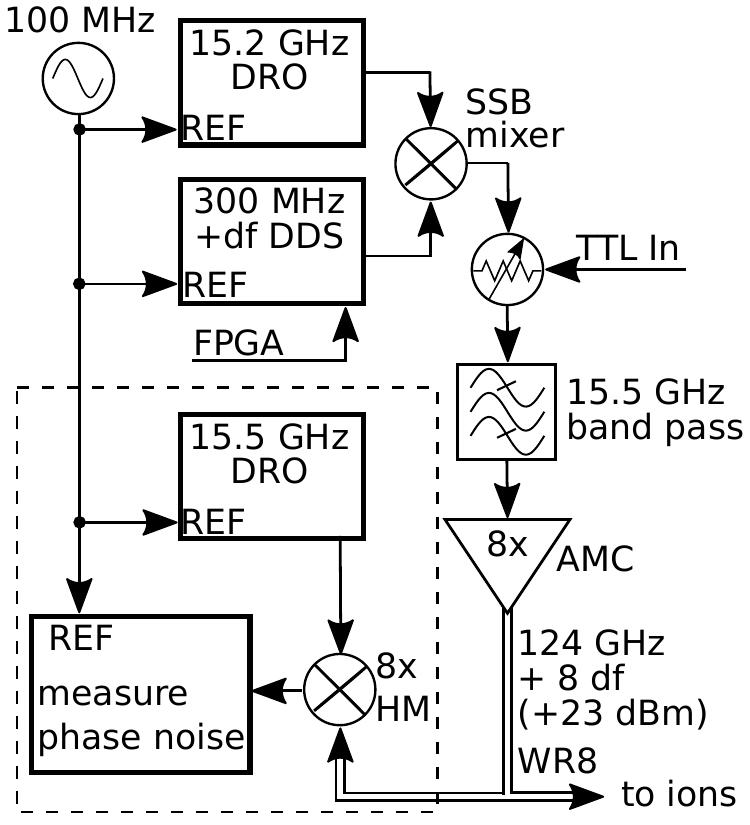}

\caption{Low phase noise synthesis chain for 124 GHz. The phase reference is
a Spectra Dynamics Inc.\cite{nist_nist} LNFR-100 low noise quartz
oscillator at $100$~MHz which is disciplined below 100~Hz by a
low phase noise $5$~MHz quartz oscillator. A 15.2~GHz DRO (Lucix
Inc.) is phase locked to the LNFR-100. Frequency and phase agility
is obtained by mixing the $15.2$~GHz with a $\sim300$~MHz tone
derived from a FPGA-controlled direct digital synthesizer (DDS, Analog
Devices AD9858) using a single-sideband (SSB) mixer (Polyphase Microwave).
Fast switching is provided by a TTL-controlled absorptive switch (Hittite).
The $15.5\,\mbox{GHz}$ sideband is filtered by a $100$~MHz passband
cavity filter (Anatech Inc.) and fed to a non-resonant, free-running
chain of room-temperature amplifiers and multipliers (AMC, Virginia
Diode Inc.). The AMC produces $\sim200\,\mbox{mW}$ at $124$~GHz.
The microwaves are transmitted to the ions in the magnet bore over
a $\sim1\,\mbox{m}$ WR8 waveguide. No horn is used; coupling to free-space
is with an open-ended WR8 waveguide. The phase noise at $124$~GHz
was measured (dashed box) on a bench top using a spectrum analyzer,
8x harmonic mixer (Millitech) and a $15.5\,\mbox{GHz}$ reference
DRO (Lucix Inc.). We observed -70~dBc at 200~Hz offset, -85~dBc
from 1~kHz to $10$~kHz offsets and -93~dBc at 100~kHz offset. }
\label{fig:124GHz_setup}
\end{figure}

Since microwave phase phase noise can also limit $T_{2}$, we use
a low-phase noise microwave synthesis chain. It starts with a low-noise
quartz oscillator at 100~MHz which is subsequently multiplied up
to 124~GHz. Due to multiplication, the phase noise at $100$~MHz
is at least $20\log_{10}(\mbox{1240})$~dB = 62~dB larger at $124$~GHz.
The ions are exposed to microwaves from outside the vacuum chamber
by a $\sim1$~m long, open-ended WR-8 waveguide. Prior to 2011, we
generated 124~GHz using a Gunn diode oscillator whose phase noise
precluded $T_{2}$ measurements beyond $\sim10$~ms \cite{biercuk_highfidelity_2009,shiga_diamagnetic_2011}.
In $2011$ the synthesis chain was upgraded as discussed in Figure~\ref{fig:124GHz_setup}
and its phase noise is not expected to impact spin coherence for $2\tau<1$~s. 

\subsection*{5. Field gradients}

Ion movement through a magnetic field gradient can produce an apparent
time-dependence to the magnetic field. Here we discuss the potential
sensitivity of our $T_{2}$ measurements to magnetic field gradients. 

Field gradients in our superconducting solenoid were minimized by
applying currents to superconducting $(z,z^{3},x,y,xz,yz,xy,x^{2}-y^{2})$
and normal $(z,z^{2})$ shim coils. Shimming the superconducting coils
was performed using a deuterium oxide ($\mbox{D}_{2}\mbox{O}$) probe
without the Penning trap inserted in the magnet bore. The field gradients
in Table~\ref{fig:gradients} were measured in $2014$ using small
crystals of ions as a field probe. The ions were translated in the
axial (radial) direction by applying a bias to an endcap (rotating
wall) electrode, and the resulting shift in electron spin-flip frequency
was measured. The measured gradients are comparable to those observed
in 2010 \cite{biercuk_highfidelity_2009}, except the linear axial
gradient, which is significantly improved. 

The gradients measured using the ions as field probe (Table~\ref{fig:gradients})
are up to an order of magnitude larger than those observed using the
$\mbox{D}_{2}\mbox{O}$ probe and otherwise empty magnet bore. This
indicates that parts of the Penning apparatus include permeable materials.
Gradient compensation with the Penning trap in the bore is possible,
but has not yet been attempted due to a non-negligible risk of magnet
quench. 

\begin{table}[!h]
\begin{tabular}{|c|c|c|}
\hline 
 & Gradients & $10^{-6}$\tabularnewline
\hline 
\hline 
$x$ & $0.42$ & $T/mm$\tabularnewline
\hline 
$y$ & $0.78$ & $T/mm$\tabularnewline
\hline 
$xy$ & $<0.2$ & $T/mm^{2}$\tabularnewline
\hline 
$x^{2}-y^{2}$ & $<0.2$ & $T/mm^{2}$\tabularnewline
\hline 
$x^{2}+y^{2}$ & $<0.1$ & $T/mm^{2}$\tabularnewline
\hline 
$z$ & $<0.03$ & $T/mm$\tabularnewline
\hline 
$z^{2}$ & $<0.2$ & $T/mm^{2}$\tabularnewline
\hline 
\end{tabular}\caption{Magnetic field inhomogeneity was quantified over a $\sim0.3\,\mbox{mm}$
radius volume using the ions' electron spin-flip as a field sensor
($28$~GHz/T). This region was sampled by applying static potentials
to endcap (rotating wall) electrodes to induce axial (radial) displacements
of the ion crystal. Room-temperature shim coils permit minimization
of the axial field gradients. }
\label{fig:gradients}
\end{table}

Ion crystal axial extent along $\hat{z}$ is sensitive to the $z$
and $z^{2}$ terms. For four planes ($\sim60\,\mu\mbox{m}$), the
axial gradients produce a spread in the electron spin-flip frequency
of $\sim50$~Hz. Ion crystal rotation averages the radial gradients
to zero, except for the $x^{2}+y^{2}$ term which can produce a $\sim170$~Hz
dispersion in the electron spin-flip frequency for a $250\,\mu\mbox{m}$
radius. The spin-echo will cancel out the effect of these gradients
as long as the ions do not move within the crystal. Although measurements
of $T_{2}$ are performed with the cooling laser blocked, previous
work showed that the period of time for a small crystal to melt is
longer than $100$~ms \cite{Jensen2004}. A further indication that
ion movement within the crystal is not limiting our $T_{2}$ measurements
is that their standard deviation is consistent with homogeneous dephasing. 

Movement of the vacuum envelope and Penning trap electrodes within
the magnet bore would result in a center-of-mass motion of the ion
crystal. This would produce a homogeneous time dependence of the magnetic
field sensed by the ions. High resolution images of the ions place
a limit on the amplitude of such motion to $<3\,\mu\mbox{m}$. A $3\,\mu\mbox{m}$
movement along the largest linear gradient ($dB/dy=780\times10^{-6}$~T/m)
generates a $\sim70$~Hz change in the spin-flip frequency. Since
we see much smaller spin-flip frequency fluctuations than $70$~Hz,
we anticipate that movement of the vacuum envelope is much less than
$3\,\mu\mbox{m}$. However, this effect may limit our $T_{2}$ coherence
times with the magnet vibrationally isolated. 

The $x-$ and $y$-linear gradients in Table~\ref{fig:gradients}
appear to be due to permeable material placed in the magnet bore,
and could be improved by reshimming the magnet with the superconducting
shim coils while using the ions to sense the gradient.

\subsection*{6. Permeable materials in magnet bore}

Permeable materials in the magnet bore give rise to additional magnetic
fields and field gradients, and their mechanical motion and temperature
variation causes time variation in the field sensed by the ions. The
Penning trap itself is constructed from low permeability materials
(type-2 titanium, Macor, aluminum, OFHC copper, Kapton and fused silica),
and the approximately cylindrical arrangement of these materials should
minimize gradients. Also in the magnet bore are relatively large structures
that guide laser beams and support photon collection optics. The structures
are not mechanically tied directly to the Penning trap. Images of
the ions constrain the axial $\hat{z}$ motion of the structures within
the magnet bore to $<3\,\mu\mbox{m}$.

The largest support structure is a $5$~cm length aluminum cylinder
($12.5$~cm OD, $7.5$~cm ID). The magnetization of a permeable
material is given by $M=\chi H=\chi B/\mu_{0}$; $\chi=2.22\times10^{-5}$
for aluminum. The field produced by a uniform magnetization can be
calculated by an equivalent current on the surface of the aluminum.
The largest gradient is at the ends of the dewar where $\frac{dBz}{dz}=630\times10^{-6}$~T/m
in the $4.5$~T field of the magnet. The fractional magnetic field
variation due to a $3\,\mu\mbox{m}$ displacement of the cylinder
is $\frac{1}{B}\frac{dB}{dz}dz=4.2\times10^{-10}$ ($52$~Hz). Field
fluctuations at the location of the ions are expected to be smaller
than this estimate by 20 to 30\%. 

The variation in the susceptibility of aluminum with temperature is
not completely negligible. We estimate a dependence of the field sensed
by the ions on the aluminum cylinder temperature to be $\frac{1}{B}\frac{dB}{dT}=3\times10^{-9}/\,^{\circ}C$.
Temperature changes will occur on time scales slower than those considered
in these experiments. 

We note that the $V_{emf}$ induced in the room temperature $Z_{0}$
shim coil was not significantly changed with the trap and supporting
structures mounted in the magnet bore. The considerations in this
section indicate that the permeable materials inserted into the magnet
bore did not significantly contribute to the $T_{2}$ measurements
with the magnet sitting on the floor. However, this effect may contribute
to $T_{2}$ coherence times with the magnet vibrationally isolated. 

\bibliographystyle{aipnum4-1}
\bibliography{paper}

\begin{thebibliography}{29}%
\makeatletter
\providecommand \@ifxundefined [1]{%
 \@ifx{#1\undefined}
}%
\providecommand \@ifnum [1]{%
 \ifnum #1\expandafter \@firstoftwo
 \else \expandafter \@secondoftwo
 \fi
}%
\providecommand \@ifx [1]{%
 \ifx #1\expandafter \@firstoftwo
 \else \expandafter \@secondoftwo
 \fi
}%
\providecommand \natexlab [1]{#1}%
\providecommand \enquote  [1]{``#1''}%
\providecommand \bibnamefont  [1]{#1}%
\providecommand \bibfnamefont [1]{#1}%
\providecommand \citenamefont [1]{#1}%
\providecommand \href@noop [0]{\@secondoftwo}%
\providecommand \href [0]{\begingroup \@sanitize@url \@href}%
\providecommand \@href[1]{\@@startlink{#1}\@@href}%
\providecommand \@@href[1]{\endgroup#1\@@endlink}%
\providecommand \@sanitize@url [0]{\catcode `\\12\catcode `\$12\catcode
  `\&12\catcode `\#12\catcode `\^12\catcode `\_12\catcode `\%12\relax}%
\providecommand \@@startlink[1]{}%
\providecommand \@@endlink[0]{}%
\providecommand \url  [0]{\begingroup\@sanitize@url \@url }%
\providecommand \@url [1]{\endgroup\@href {#1}{\urlprefix }}%
\providecommand \urlprefix  [0]{URL }%
\providecommand \Eprint [0]{\href }%
\providecommand \doibase [0]{http://dx.doi.org/}%
\providecommand \selectlanguage [0]{\@gobble}%
\providecommand \bibinfo  [0]{\@secondoftwo}%
\providecommand \bibfield  [0]{\@secondoftwo}%
\providecommand \translation [1]{[#1]}%
\providecommand \BibitemOpen [0]{}%
\providecommand \bibitemStop [0]{}%
\providecommand \bibitemNoStop [0]{.\EOS\space}%
\providecommand \EOS [0]{\spacefactor3000\relax}%
\providecommand \BibitemShut  [1]{\csname bibitem#1\endcsname}%
\let\auto@bib@innerbib\@empty
\bibitem [{\citenamefont {Van~Dyck}\ \emph {et~al.}(1999)\citenamefont
  {Van~Dyck}, \citenamefont {Farnham}, \citenamefont {Zafonte},\ and\
  \citenamefont {Schwinberg}}]{vandyck_ultrastable_1999}%
  \BibitemOpen
  \bibfield  {author} {\bibinfo {author} {\bibfnamefont {R.~S.}\ \bibnamefont
  {Van~Dyck}}, \bibinfo {author} {\bibfnamefont {D.~L.}\ \bibnamefont
  {Farnham}}, \bibinfo {author} {\bibfnamefont {S.~L.}\ \bibnamefont
  {Zafonte}}, \ and\ \bibinfo {author} {\bibfnamefont {P.~B.}\ \bibnamefont
  {Schwinberg}},\ }\href@noop {} {\bibfield  {journal} {\bibinfo  {journal}
  {Rev. Sci. Instrum.}\ }\textbf {\bibinfo {volume} {70}},\ \bibinfo {pages}
  {1665} (\bibinfo {year} {1999})}\BibitemShut {NoStop}%
\bibitem [{\citenamefont {Myers}(2013)}]{myers_most_2013}%
  \BibitemOpen
  \bibfield  {author} {\bibinfo {author} {\bibfnamefont {E.~G.}\ \bibnamefont
  {Myers}},\ }\href@noop {} {\bibfield  {journal} {\bibinfo  {journal} {Int. J.
  Mass Spectrom.}\ }\textbf {\bibinfo {volume} {349-350}},\ \bibinfo {pages}
  {107} (\bibinfo {year} {2013})}\BibitemShut {NoStop}%
\bibitem [{\citenamefont {Rainville}, \citenamefont {Thompson},\ and\
  \citenamefont {Pritchard}(2004)}]{rainville_ion_2004}%
  \BibitemOpen
  \bibfield  {author} {\bibinfo {author} {\bibfnamefont {S.}~\bibnamefont
  {Rainville}}, \bibinfo {author} {\bibfnamefont {J.~K.}\ \bibnamefont
  {Thompson}}, \ and\ \bibinfo {author} {\bibfnamefont {D.~E.}\ \bibnamefont
  {Pritchard}},\ }\href@noop {} {\bibfield  {journal} {\bibinfo  {journal}
  {Science}\ }\textbf {\bibinfo {volume} {303}},\ \bibinfo {pages} {334}
  (\bibinfo {year} {2004})}\BibitemShut {NoStop}%
\bibitem [{\citenamefont {Odom}\ \emph {et~al.}(2006)\citenamefont {Odom},
  \citenamefont {Hanneke}, \citenamefont {{B. D'Urso}},\ and\ \citenamefont
  {Gabrielse}}]{odom_new_2006}%
  \BibitemOpen
  \bibfield  {author} {\bibinfo {author} {\bibfnamefont {B.}~\bibnamefont
  {Odom}}, \bibinfo {author} {\bibfnamefont {D.}~\bibnamefont {Hanneke}},
  \bibinfo {author} {\bibnamefont {{B. D'Urso}}}, \ and\ \bibinfo {author}
  {\bibfnamefont {G.}~\bibnamefont {Gabrielse}},\ }\href@noop {} {\bibfield
  {journal} {\bibinfo  {journal} {Phys. Rev. Lett.}\ }\textbf {\bibinfo
  {volume} {97}} (\bibinfo {year} {2006})}\BibitemShut {NoStop}%
\bibitem [{\citenamefont {Hanneke}, \citenamefont {Fogwell},\ and\
  \citenamefont {Gabrielse}(2008)}]{hanneke_new_2008}%
  \BibitemOpen
  \bibfield  {author} {\bibinfo {author} {\bibfnamefont {D.}~\bibnamefont
  {Hanneke}}, \bibinfo {author} {\bibfnamefont {S.}~\bibnamefont {Fogwell}}, \
  and\ \bibinfo {author} {\bibfnamefont {G.}~\bibnamefont {Gabrielse}},\
  }\href@noop {} {\bibfield  {journal} {\bibinfo  {journal} {Phys. Rev. Lett.}\
  }\textbf {\bibinfo {volume} {100}} (\bibinfo {year} {2008})}\BibitemShut
  {NoStop}%
\bibitem [{\citenamefont {Sturm}\ \emph {et~al.}(2011)\citenamefont {Sturm},
  \citenamefont {Wagner}, \citenamefont {Schabinger}, \citenamefont {Zatorski},
  \citenamefont {Harman}, \citenamefont {Quint}, \citenamefont {Werth},
  \citenamefont {Keitel},\ and\ \citenamefont {Blaum}}]{sturm_g_2011}%
  \BibitemOpen
  \bibfield  {author} {\bibinfo {author} {\bibfnamefont {S.}~\bibnamefont
  {Sturm}}, \bibinfo {author} {\bibfnamefont {A.}~\bibnamefont {Wagner}},
  \bibinfo {author} {\bibfnamefont {B.}~\bibnamefont {Schabinger}}, \bibinfo
  {author} {\bibfnamefont {J.}~\bibnamefont {Zatorski}}, \bibinfo {author}
  {\bibfnamefont {Z.}~\bibnamefont {Harman}}, \bibinfo {author} {\bibfnamefont
  {W.}~\bibnamefont {Quint}}, \bibinfo {author} {\bibfnamefont
  {G.}~\bibnamefont {Werth}}, \bibinfo {author} {\bibfnamefont
  {C.}~\bibnamefont {Keitel}}, \ and\ \bibinfo {author} {\bibfnamefont
  {K.}~\bibnamefont {Blaum}},\ }\href@noop {} {\bibfield  {journal} {\bibinfo
  {journal} {Phys. Rev. Lett.}\ }\textbf {\bibinfo {volume} {107}} (\bibinfo
  {year} {2011})}\BibitemShut {NoStop}%
\bibitem [{\citenamefont {Sturm}\ \emph {et~al.}(2014)\citenamefont {Sturm},
  \citenamefont {K{\"o}hler}, \citenamefont {Zatorski}, \citenamefont {Wagner},
  \citenamefont {Harman}, \citenamefont {Werth}, \citenamefont {Quint},
  \citenamefont {Keitel},\ and\ \citenamefont
  {Blaum}}]{sturm_highprecision_2014}%
  \BibitemOpen
  \bibfield  {author} {\bibinfo {author} {\bibfnamefont {S.}~\bibnamefont
  {Sturm}}, \bibinfo {author} {\bibfnamefont {F.}~\bibnamefont {K{\"o}hler}},
  \bibinfo {author} {\bibfnamefont {J.}~\bibnamefont {Zatorski}}, \bibinfo
  {author} {\bibfnamefont {A.}~\bibnamefont {Wagner}}, \bibinfo {author}
  {\bibfnamefont {Z.}~\bibnamefont {Harman}}, \bibinfo {author} {\bibfnamefont
  {G.}~\bibnamefont {Werth}}, \bibinfo {author} {\bibfnamefont
  {W.}~\bibnamefont {Quint}}, \bibinfo {author} {\bibfnamefont {C.~H.}\
  \bibnamefont {Keitel}}, \ and\ \bibinfo {author} {\bibfnamefont
  {K.}~\bibnamefont {Blaum}},\ }\href@noop {} {\bibfield  {journal} {\bibinfo
  {journal} {Nature}\ }\textbf {\bibinfo {volume} {506}},\ \bibinfo {pages}
  {467} (\bibinfo {year} {2014})}\BibitemShut {NoStop}%
\bibitem [{\citenamefont {Salinger}\ and\ \citenamefont
  {Wheatley}(1961)}]{salinger_magnetic_1961}%
  \BibitemOpen
  \bibfield  {author} {\bibinfo {author} {\bibfnamefont {G.~L.}\ \bibnamefont
  {Salinger}}\ and\ \bibinfo {author} {\bibfnamefont {J.~C.}\ \bibnamefont
  {Wheatley}},\ }\href@noop {} {\bibfield  {journal} {\bibinfo  {journal} {Rev.
  Sci. Instrum.}\ }\textbf {\bibinfo {volume} {32}},\ \bibinfo {pages} {872}
  (\bibinfo {year} {1961})}\BibitemShut {NoStop}%
\bibitem [{\citenamefont {Xia}\ \emph {et~al.}(2002)\citenamefont {Xia},
  \citenamefont {Bray-Ali}, \citenamefont {Zhang}, \citenamefont {Fink},
  \citenamefont {White}, \citenamefont {Gould},\ and\ \citenamefont
  {Bozler}}]{xia_magnetization_2002}%
  \BibitemOpen
  \bibfield  {author} {\bibinfo {author} {\bibfnamefont {Z.}~\bibnamefont
  {Xia}}, \bibinfo {author} {\bibfnamefont {J.}~\bibnamefont {Bray-Ali}},
  \bibinfo {author} {\bibfnamefont {J.}~\bibnamefont {Zhang}}, \bibinfo
  {author} {\bibfnamefont {R.~B.}\ \bibnamefont {Fink}}, \bibinfo {author}
  {\bibfnamefont {K.~S.}\ \bibnamefont {White}}, \bibinfo {author}
  {\bibfnamefont {C.~M.}\ \bibnamefont {Gould}}, \ and\ \bibinfo {author}
  {\bibfnamefont {H.~M.}\ \bibnamefont {Bozler}},\ }\href@noop {} {\bibfield
  {journal} {\bibinfo  {journal} {J. Low Temp. Phys.}\ }\textbf {\bibinfo
  {volume} {126}},\ \bibinfo {pages} {655} (\bibinfo {year}
  {2002})}\BibitemShut {NoStop}%
\bibitem [{\citenamefont {Gabrielse}\ and\ \citenamefont
  {Tan}(1988)}]{gabrielse_selfshielding_1988}%
  \BibitemOpen
  \bibfield  {author} {\bibinfo {author} {\bibfnamefont {G.}~\bibnamefont
  {Gabrielse}}\ and\ \bibinfo {author} {\bibfnamefont {J.}~\bibnamefont
  {Tan}},\ }\href@noop {} {\bibfield  {journal} {\bibinfo  {journal} {J. Appl.
  Phys.}\ }\textbf {\bibinfo {volume} {63}},\ \bibinfo {pages} {5143} (\bibinfo
  {year} {1988})}\BibitemShut {NoStop}%
\bibitem [{\citenamefont
  {Schneider-Muntau}(1997)}]{schneider-mantau1997high_field_magnets}%
  \BibitemOpen
  \bibfield  {author} {\bibinfo {author} {\bibfnamefont {H.~J.}\ \bibnamefont
  {Schneider-Muntau}},\ }\href@noop {} {\bibfield  {journal} {\bibinfo
  {journal} {Solid State Nuclear Magnetic Resonance}\ }\textbf {\bibinfo
  {volume} {9}},\ \bibinfo {pages} {61} (\bibinfo {year} {1997})}\BibitemShut
  {NoStop}%
\bibitem [{\citenamefont {Kiyoshi}\ \emph {et~al.}(2010)\citenamefont
  {Kiyoshi}, \citenamefont {{Seyong Choi}}, \citenamefont {Matsumoto},
  \citenamefont {Zaitsu}, \citenamefont {Hase}, \citenamefont {Miyazaki},
  \citenamefont {Otsuka}, \citenamefont {Yoshikawa}, \citenamefont {Hamada},
  \citenamefont {Hosono}, \citenamefont {Yanagisawa}, \citenamefont {Nakagome},
  \citenamefont {Takahashi}, \citenamefont {Yamazaki},\ and\ \citenamefont
  {Maeda}}]{kiyoshi2010hts_nmr}%
  \BibitemOpen
  \bibfield  {author} {\bibinfo {author} {\bibfnamefont {T.}~\bibnamefont
  {Kiyoshi}}, \bibinfo {author} {\bibnamefont {{Seyong Choi}}}, \bibinfo
  {author} {\bibfnamefont {S.}~\bibnamefont {Matsumoto}}, \bibinfo {author}
  {\bibfnamefont {K.}~\bibnamefont {Zaitsu}}, \bibinfo {author} {\bibfnamefont
  {T.}~\bibnamefont {Hase}}, \bibinfo {author} {\bibfnamefont {T.}~\bibnamefont
  {Miyazaki}}, \bibinfo {author} {\bibfnamefont {A.}~\bibnamefont {Otsuka}},
  \bibinfo {author} {\bibfnamefont {M.}~\bibnamefont {Yoshikawa}}, \bibinfo
  {author} {\bibfnamefont {M.}~\bibnamefont {Hamada}}, \bibinfo {author}
  {\bibfnamefont {M.}~\bibnamefont {Hosono}}, \bibinfo {author} {\bibfnamefont
  {Y.}~\bibnamefont {Yanagisawa}}, \bibinfo {author} {\bibfnamefont
  {H.}~\bibnamefont {Nakagome}}, \bibinfo {author} {\bibfnamefont
  {M.}~\bibnamefont {Takahashi}}, \bibinfo {author} {\bibfnamefont
  {T.}~\bibnamefont {Yamazaki}}, \ and\ \bibinfo {author} {\bibfnamefont
  {H.}~\bibnamefont {Maeda}},\ }\href {\doibase 10.1109/TASC.2010.2040148}
  {\bibfield  {journal} {\bibinfo  {journal} {IEEE Transactions on Applied
  Superconductivity}\ }\textbf {\bibinfo {volume} {20}},\ \bibinfo {pages}
  {714} (\bibinfo {year} {2010})}\BibitemShut {NoStop}%
\bibitem [{\citenamefont {Biercuk}\ \emph
  {et~al.}(2009{\natexlab{a}})\citenamefont {Biercuk}, \citenamefont {Uys},
  \citenamefont {{VanDevender}}, \citenamefont {Shiga}, \citenamefont {Itano},\
  and\ \citenamefont {Bollinger}}]{Biercuk2009}%
  \BibitemOpen
  \bibfield  {author} {\bibinfo {author} {\bibfnamefont {M.~J.}\ \bibnamefont
  {Biercuk}}, \bibinfo {author} {\bibfnamefont {H.}~\bibnamefont {Uys}},
  \bibinfo {author} {\bibfnamefont {A.~P.}\ \bibnamefont {{VanDevender}}},
  \bibinfo {author} {\bibfnamefont {N.}~\bibnamefont {Shiga}}, \bibinfo
  {author} {\bibfnamefont {W.~M.}\ \bibnamefont {Itano}}, \ and\ \bibinfo
  {author} {\bibfnamefont {J.~J.}\ \bibnamefont {Bollinger}},\ }\href@noop {}
  {\bibfield  {journal} {\bibinfo  {journal} {Nature}\ }\textbf {\bibinfo
  {volume} {458}},\ \bibinfo {pages} {996} (\bibinfo {year}
  {2009}{\natexlab{a}})}\BibitemShut {NoStop}%
\bibitem [{\citenamefont {Marshall}, \citenamefont {Hendrickson},\ and\
  \citenamefont {Jackson}(1998)}]{marshall_fourier_1998}%
  \BibitemOpen
  \bibfield  {author} {\bibinfo {author} {\bibfnamefont {A.~G.}\ \bibnamefont
  {Marshall}}, \bibinfo {author} {\bibfnamefont {C.~L.}\ \bibnamefont
  {Hendrickson}}, \ and\ \bibinfo {author} {\bibfnamefont {G.~S.}\ \bibnamefont
  {Jackson}},\ }\href@noop {} {\bibfield  {journal} {\bibinfo  {journal} {Mass
  Spectrom. Rev.}\ }\textbf {\bibinfo {volume} {17}},\ \bibinfo {pages} {1}
  (\bibinfo {year} {1998})}\BibitemShut {NoStop}%
\bibitem [{\citenamefont {Britton}\ \emph {et~al.}(2012)\citenamefont
  {Britton}, \citenamefont {Sawyer}, \citenamefont {Keith}, \citenamefont
  {Wang}, \citenamefont {Freericks}, \citenamefont {Uys}, \citenamefont
  {Biercuk},\ and\ \citenamefont {Bollinger}}]{britton_engineered_2012}%
  \BibitemOpen
  \bibfield  {author} {\bibinfo {author} {\bibfnamefont {J.~W.}\ \bibnamefont
  {Britton}}, \bibinfo {author} {\bibfnamefont {B.~C.}\ \bibnamefont {Sawyer}},
  \bibinfo {author} {\bibfnamefont {A.~C.}\ \bibnamefont {Keith}}, \bibinfo
  {author} {\bibfnamefont {C.-C.~J.}\ \bibnamefont {Wang}}, \bibinfo {author}
  {\bibfnamefont {J.~K.}\ \bibnamefont {Freericks}}, \bibinfo {author}
  {\bibfnamefont {H.}~\bibnamefont {Uys}}, \bibinfo {author} {\bibfnamefont
  {M.~J.}\ \bibnamefont {Biercuk}}, \ and\ \bibinfo {author} {\bibfnamefont
  {J.~J.}\ \bibnamefont {Bollinger}},\ }\href@noop {} {\bibfield  {journal}
  {\bibinfo  {journal} {Nature}\ }\textbf {\bibinfo {volume} {484}},\ \bibinfo
  {pages} {489} (\bibinfo {year} {2012})}\BibitemShut {NoStop}%
\bibitem [{\citenamefont {Porras}\ and\ \citenamefont
  {{Cirac}}(2006)}]{Porras2006}%
  \BibitemOpen
  \bibfield  {author} {\bibinfo {author} {\bibfnamefont {D.}~\bibnamefont
  {Porras}}\ and\ \bibinfo {author} {\bibfnamefont {J.}~\bibnamefont
  {{Cirac}}},\ }\href@noop {} {\bibfield  {journal} {\bibinfo  {journal} {Phys.
  {Rev}. {Lett}.}\ }\textbf {\bibinfo {volume} {96}},\ \bibinfo {pages}
  {250501} (\bibinfo {year} {2006})}\BibitemShut {NoStop}%
\bibitem [{\citenamefont {Baltrusch}\ \emph {et~al.}(2011)\citenamefont
  {Baltrusch}, \citenamefont {Negretti}, \citenamefont {Taylor},\ and\
  \citenamefont {Calarco}}]{baltrusch_fast_2011}%
  \BibitemOpen
  \bibfield  {author} {\bibinfo {author} {\bibfnamefont {J.}~\bibnamefont
  {Baltrusch}}, \bibinfo {author} {\bibfnamefont {A.}~\bibnamefont {Negretti}},
  \bibinfo {author} {\bibfnamefont {J.}~\bibnamefont {Taylor}}, \ and\ \bibinfo
  {author} {\bibfnamefont {T.}~\bibnamefont {Calarco}},\ }\href@noop {}
  {\bibfield  {journal} {\bibinfo  {journal} {Phys. Rev. A}\ }\textbf {\bibinfo
  {volume} {83}},\ \bibinfo {pages} {042319} (\bibinfo {year}
  {2011})}\BibitemShut {NoStop}%
\bibitem [{\citenamefont {Biercuk}\ \emph
  {et~al.}(2009{\natexlab{b}})\citenamefont {Biercuk}, \citenamefont {Uys},
  \citenamefont {VanDevender}, \citenamefont {Shiga}, \citenamefont {Itano},\
  and\ \citenamefont {Bollinger}}]{biercuk_highfidelity_2009}%
  \BibitemOpen
  \bibfield  {author} {\bibinfo {author} {\bibfnamefont {M.~J.}\ \bibnamefont
  {Biercuk}}, \bibinfo {author} {\bibfnamefont {H.}~\bibnamefont {Uys}},
  \bibinfo {author} {\bibfnamefont {A.~P.}\ \bibnamefont {VanDevender}},
  \bibinfo {author} {\bibfnamefont {N.}~\bibnamefont {Shiga}}, \bibinfo
  {author} {\bibfnamefont {W.~M.}\ \bibnamefont {Itano}}, \ and\ \bibinfo
  {author} {\bibfnamefont {J.~J.}\ \bibnamefont {Bollinger}},\ }\href@noop {}
  {\bibfield  {journal} {\bibinfo  {journal} {Quantum Inf. Comput.}\ ,\
  \bibinfo {pages} {920}} (\bibinfo {year} {2009}{\natexlab{b}})}\BibitemShut
  {NoStop}%
\bibitem [{nis()}]{nist_nist}%
  \BibitemOpen
  \href@noop {} {\enquote {\bibinfo {title} {{NIST does not endorse commercial
  products. We provide reference to the suppliers of the specific
  instrumentation used in this work for informational purposes only.}}}\
  }\BibitemShut {NoStop}%
\bibitem [{\citenamefont {{Morley}}, \citenamefont {{Brunel}},\ and\
  \citenamefont {{van Tol}}(2008)}]{morley_multifrequency_2008}%
  \BibitemOpen
  \bibfield  {author} {\bibinfo {author} {\bibfnamefont {G.~W.}\ \bibnamefont
  {{Morley}}}, \bibinfo {author} {\bibfnamefont {L.}~\bibnamefont {{Brunel}}},
  \ and\ \bibinfo {author} {\bibfnamefont {J.}~\bibnamefont {{van Tol}}},\
  }\href@noop {} {\bibfield  {journal} {\bibinfo  {journal} {Rev. {Sci}.
  {Instrum}.}\ }\textbf {\bibinfo {volume} {79}},\ \bibinfo {pages} {064703}
  (\bibinfo {year} {2008})}\BibitemShut {NoStop}%
\bibitem [{\citenamefont {Biercuk}, \citenamefont {Doherty},\ and\
  \citenamefont {Uys}(2011)}]{biercuk_dynamical_2011}%
  \BibitemOpen
  \bibfield  {author} {\bibinfo {author} {\bibfnamefont {M.~J.}\ \bibnamefont
  {Biercuk}}, \bibinfo {author} {\bibfnamefont {A.~C.}\ \bibnamefont
  {Doherty}}, \ and\ \bibinfo {author} {\bibfnamefont {H.}~\bibnamefont
  {Uys}},\ }\href@noop {} {\bibfield  {journal} {\bibinfo  {journal} {J. Phys.
  B}\ }\textbf {\bibinfo {volume} {44}},\ \bibinfo {pages} {154002} (\bibinfo
  {year} {2011})}\BibitemShut {NoStop}%
\bibitem [{\citenamefont {Ahokas}\ \emph {et~al.}(2008)\citenamefont {Ahokas},
  \citenamefont {{J. Jarvinen}}, \citenamefont {Shlyapnikov},\ and\
  \citenamefont {{S. Vasiliev}}}]{ahokas_clock_2008}%
  \BibitemOpen
  \bibfield  {author} {\bibinfo {author} {\bibfnamefont {J.}~\bibnamefont
  {Ahokas}}, \bibinfo {author} {\bibnamefont {{J. Jarvinen}}}, \bibinfo
  {author} {\bibfnamefont {G.}~\bibnamefont {Shlyapnikov}}, \ and\ \bibinfo
  {author} {\bibnamefont {{S. Vasiliev}}},\ }\href@noop {} {\bibfield
  {journal} {\bibinfo  {journal} {Phys. Rev. Lett.}\ }\textbf {\bibinfo
  {volume} {101}} (\bibinfo {year} {2008})}\BibitemShut {NoStop}%
\bibitem [{\citenamefont {Takahashi}\ \emph {et~al.}(2008)\citenamefont
  {Takahashi}, \citenamefont {Hanson}, \citenamefont {{van Tol}}, \citenamefont
  {Sherwin},\ and\ \citenamefont {Awschalom}}]{takahashi_quenching_2008}%
  \BibitemOpen
  \bibfield  {author} {\bibinfo {author} {\bibfnamefont {S.}~\bibnamefont
  {Takahashi}}, \bibinfo {author} {\bibfnamefont {R.}~\bibnamefont {Hanson}},
  \bibinfo {author} {\bibfnamefont {J.}~\bibnamefont {{van Tol}}}, \bibinfo
  {author} {\bibfnamefont {M.}~\bibnamefont {Sherwin}}, \ and\ \bibinfo
  {author} {\bibfnamefont {D.}~\bibnamefont {Awschalom}},\ }\href@noop {}
  {\bibfield  {journal} {\bibinfo  {journal} {Phys. Rev. Lett.}\ }\textbf
  {\bibinfo {volume} {101}} (\bibinfo {year} {2008})}\BibitemShut {NoStop}%
\bibitem [{\citenamefont {Morley}\ \emph {et~al.}(2008)\citenamefont {Morley},
  \citenamefont {{McCamey}}, \citenamefont {Seipel}, \citenamefont {Brunel},
  \citenamefont {{van Tol}},\ and\ \citenamefont
  {Boehme}}]{morley_longlived_2008}%
  \BibitemOpen
  \bibfield  {author} {\bibinfo {author} {\bibfnamefont {G.}~\bibnamefont
  {Morley}}, \bibinfo {author} {\bibfnamefont {D.}~\bibnamefont {{McCamey}}},
  \bibinfo {author} {\bibfnamefont {H.}~\bibnamefont {Seipel}}, \bibinfo
  {author} {\bibfnamefont {L.-C.}\ \bibnamefont {Brunel}}, \bibinfo {author}
  {\bibfnamefont {J.}~\bibnamefont {{van Tol}}}, \ and\ \bibinfo {author}
  {\bibfnamefont {C.}~\bibnamefont {Boehme}},\ }\href@noop {} {\bibfield
  {journal} {\bibinfo  {journal} {Phys. Rev. Lett.}\ }\textbf {\bibinfo
  {volume} {101}} (\bibinfo {year} {2008})}\BibitemShut {NoStop}%
\bibitem [{\citenamefont {Tyryshkin}\ \emph {et~al.}(2011)\citenamefont
  {Tyryshkin}, \citenamefont {Tojo}, \citenamefont {Morton}, \citenamefont
  {Riemann}, \citenamefont {Abrosimov}, \citenamefont {Becker}, \citenamefont
  {Pohl}, \citenamefont {Schenkel}, \citenamefont {Thewalt}, \citenamefont
  {Itoh},\ and\ \citenamefont {Lyon}}]{tyryshkin_electron_2011}%
  \BibitemOpen
  \bibfield  {author} {\bibinfo {author} {\bibfnamefont {A.~M.}\ \bibnamefont
  {Tyryshkin}}, \bibinfo {author} {\bibfnamefont {S.}~\bibnamefont {Tojo}},
  \bibinfo {author} {\bibfnamefont {J.~J.~L.}\ \bibnamefont {Morton}}, \bibinfo
  {author} {\bibfnamefont {H.}~\bibnamefont {Riemann}}, \bibinfo {author}
  {\bibfnamefont {N.~V.}\ \bibnamefont {Abrosimov}}, \bibinfo {author}
  {\bibfnamefont {P.}~\bibnamefont {Becker}}, \bibinfo {author} {\bibfnamefont
  {H.-J.}\ \bibnamefont {Pohl}}, \bibinfo {author} {\bibfnamefont
  {T.}~\bibnamefont {Schenkel}}, \bibinfo {author} {\bibfnamefont {M.~L.~W.}\
  \bibnamefont {Thewalt}}, \bibinfo {author} {\bibfnamefont {K.~M.}\
  \bibnamefont {Itoh}}, \ and\ \bibinfo {author} {\bibfnamefont {S.~A.}\
  \bibnamefont {Lyon}},\ }\href@noop {} {\bibfield  {journal} {\bibinfo
  {journal} {Nat. Mater.}\ }\textbf {\bibinfo {volume} {11}},\ \bibinfo {pages}
  {143} (\bibinfo {year} {2011})}\BibitemShut {NoStop}%
\bibitem [{\citenamefont {Mitchell}\ \emph {et~al.}(1998)\citenamefont
  {Mitchell}, \citenamefont {Bollinger}, \citenamefont {Dubin}, \citenamefont
  {Huang}, \citenamefont {Itano},\ and\ \citenamefont
  {Baughman}}]{mitchell_direct_1998}%
  \BibitemOpen
  \bibfield  {author} {\bibinfo {author} {\bibfnamefont {T.~B.}\ \bibnamefont
  {Mitchell}}, \bibinfo {author} {\bibfnamefont {J.~J.}\ \bibnamefont
  {Bollinger}}, \bibinfo {author} {\bibfnamefont {D.}~\bibnamefont {Dubin}},
  \bibinfo {author} {\bibfnamefont {X.}~\bibnamefont {Huang}}, \bibinfo
  {author} {\bibfnamefont {W.~M.}\ \bibnamefont {Itano}}, \ and\ \bibinfo
  {author} {\bibfnamefont {R.~H.}\ \bibnamefont {Baughman}},\ }\href@noop {}
  {\bibfield  {journal} {\bibinfo  {journal} {Science}\ }\textbf {\bibinfo
  {volume} {282}},\ \bibinfo {pages} {1290} (\bibinfo {year}
  {1998})}\BibitemShut {NoStop}%
\bibitem [{\citenamefont {Sawyer}\ \emph {et~al.}(2012)\citenamefont {Sawyer},
  \citenamefont {{Britton}}, \citenamefont {{Keith}}, \citenamefont {{Wang}},
  \citenamefont {{Freericks}}, \citenamefont {{Uys}}, \citenamefont
  {{Biercuk}},\ and\ \citenamefont {{Bollinger}}}]{sawyer_spectroscopy_2012}%
  \BibitemOpen
  \bibfield  {author} {\bibinfo {author} {\bibfnamefont {B.~C.}\ \bibnamefont
  {Sawyer}}, \bibinfo {author} {\bibfnamefont {J.~W.}\ \bibnamefont
  {{Britton}}}, \bibinfo {author} {\bibfnamefont {A.~C.}\ \bibnamefont
  {{Keith}}}, \bibinfo {author} {\bibfnamefont {C.-C.~J.}\ \bibnamefont
  {{Wang}}}, \bibinfo {author} {\bibfnamefont {J.~K.}\ \bibnamefont
  {{Freericks}}}, \bibinfo {author} {\bibfnamefont {H.}~\bibnamefont {{Uys}}},
  \bibinfo {author} {\bibfnamefont {M.~J.}\ \bibnamefont {{Biercuk}}}, \ and\
  \bibinfo {author} {\bibfnamefont {J.~J.}\ \bibnamefont {{Bollinger}}},\
  }\href@noop {} {\bibfield  {journal} {\bibinfo  {journal} {Phys. {Rev}.
  {Lett}.}\ }\textbf {\bibinfo {volume} {108}} (\bibinfo {year}
  {2012})}\BibitemShut {NoStop}%
\bibitem [{\citenamefont {Shiga}, \citenamefont {{Itano}},\ and\ \citenamefont
  {{Bollinger}}(2011)}]{shiga_diamagnetic_2011}%
  \BibitemOpen
  \bibfield  {author} {\bibinfo {author} {\bibfnamefont {N.}~\bibnamefont
  {Shiga}}, \bibinfo {author} {\bibfnamefont {W.~M.}\ \bibnamefont {{Itano}}},
  \ and\ \bibinfo {author} {\bibfnamefont {J.~J.}\ \bibnamefont
  {{Bollinger}}},\ }\href@noop {} {\bibfield  {journal} {\bibinfo  {journal}
  {Phys. {Rev}. {A}}\ }\textbf {\bibinfo {volume} {84}} (\bibinfo {year}
  {2011})}\BibitemShut {NoStop}%
\bibitem [{\citenamefont {Jensen}, \citenamefont {Hasegawa},\ and\
  \citenamefont {Bollinger}(2004)}]{Jensen2004}%
  \BibitemOpen
  \bibfield  {author} {\bibinfo {author} {\bibfnamefont {M.}~\bibnamefont
  {Jensen}}, \bibinfo {author} {\bibfnamefont {T.}~\bibnamefont {Hasegawa}}, \
  and\ \bibinfo {author} {\bibfnamefont {J.~J.}\ \bibnamefont {Bollinger}},\
  }\href@noop {} {\bibfield  {journal} {\bibinfo  {journal} {Phys. Rev. A}\
  }\textbf {\bibinfo {volume} {70}} (\bibinfo {year} {2004})}\BibitemShut
  {NoStop}%
\end{thebibliography}%

\end{document}